\documentclass[aps,prb,twocolumn,superscriptaddress,showpacs]{revtex4-1}

\usepackage{graphicx}
\usepackage{amsmath}
\usepackage{longtable}
\usepackage{color}

\begin{document}

\title{Chemical tuning between triangular and honeycomb structures\\ in a 5\emph{d} spin-orbit Mott insulator}

\author{R. D. Johnson}
\email{roger.johnson@physics.ox.ac.uk}
\affiliation{Clarendon Laboratory, Department of Physics, University of Oxford, Oxford, OX1 3PU, United Kingdom}
\author{I. Broeders}
\affiliation{Clarendon Laboratory, Department of Physics, University of Oxford, Oxford, OX1 3PU, United Kingdom}
\author{K. Mehlawat}
\altaffiliation[Current address: ]{IFW Dresden, Helmholtzstraße 20, D-01069 Dresden, Germany}
\affiliation{Department of Physical Sciences, Indian Institute of Science Education and Research, Knowledge City, Sector 81, Mohali 140306, India}
\author{Y. Li}
\affiliation{Institut f\"{u}r Theoretische Physik,
Goethe-Universit\"{a}t Frankfurt, 60438 Frankfurt am Main,
Germany}
\affiliation{Department of Applied Physics and MOE Key Laboratory for Nonequilibrium Synthesis and Modulation of Condensed Matter, School of Science,
Xi'an Jiaotong University, Xi'an 710049, China}
\author{Y. Singh}
\affiliation{Department of Physical Sciences, Indian Institute of Science Education and Research, Knowledge City, Sector 81, Mohali 140306, India}
\author{R. Valent{\'\i}}
\affiliation{Institut f\"{u}r Theoretische Physik,
Goethe-Universit\"{a}t Frankfurt, 60438 Frankfurt am Main,
Germany}
\author{R. Coldea}
\affiliation{Clarendon Laboratory, Department of Physics, University of Oxford, Oxford, OX1 3PU, United Kingdom}

\date{\today}

\begin{abstract}

We report structural studies of the spin-orbit Mott insulator family
	K$_x$Ir$_y$O$_2$, with triangular layers of edge-sharing IrO$_6$
	octahedra bonded by potassium ions. The potassium content acts as a
	chemical tuning parameter to control the amount of charge in the Ir-O
	layers. Unlike the isostructural families with Ir replaced by Co or Rh
	($y=1$), which are metallic over a  range of potassium compositions
	$x$, we instead find insulating behaviour with charge neutrality
	achieved via iridium vacancies, which order in a honeycomb supercell
	above a critical composition $x_c$. By performing density functional theory calculations
	we attribute the observed behaviour
	to a subtle interplay of crystal-field environment, local electronic correlations
	and strong spin-orbit interaction at the Ir$^{4+}$ sites, making
	this structural family a candidate to display Kitaev magnetism in the
	experimentally unexplored regime that interpolates between triangular
	and honeycomb structures.
\end{abstract}

\maketitle

\section{Introduction}

Mott insulators with strong spin-orbit interactions, as realized in 5$d$ oxides, are attracting much interest as candidates to display novel cooperative magnetic behaviours such as quantum spin liquids with exotic quasiparticles, anisotropic spin fluctuations with momentum-dependent polarization, and unconventional magnetic orders such as counter-rotating incommensurate antiferromagnetism (for recent reviews see references \citenum{rev1} and \citenum{rev2}). Such physics is stabilized by strong frustration effects  from bond-dependent anisotropic (Kitaev) interactions between spin-orbit entangled magnetic moments. Experimental evidence for such interactions has been found in the layered honeycomb materials $\alpha$-Li$_2$IrO$_3$ \cite{Williams}, Na$_2$IrO$_3$ \cite{Chun15}, $\alpha$-RuCl$_3$ \cite{Banerjee15}, and the three-dimensional harmonic honeycombs $\beta$- and $\gamma$-Li$_2$IrO$_3$ \cite{Takayama,BiffinPRB,BiffinPRL}. Rich physics stabilized by Kitaev interactions is also expected on other lattice geometries, in particular on the 2D triangular structure incommensurate magnetic vortex crystals and spin nematic orders have been theoretically predicted \cite{Rousochatzakis,Kimchi2014}, but are yet to be experimentally realized.
Notable examples of triangular structures composed of edge-sharing octahedra (\emph{i.e.} a similar transition-metal bonding environment to the above honeycomb structures) are Na$_x$CoO$_2$ \cite{NaxCoO2}, K$_x$CoO$_2$ \cite{Jansen,Hironaka,Nakamura} and K$_x$RhO$_2$ \cite{Shibasaki_2010,Zhang}. These materials show metallic behaviour for a range of compositions, $x$, with the transition metal adopting a non-integer valence state $+(4-x)$. Here, we report a structural study of the related family, K$_x$Ir$_y$O$_2$, which in contrast show insulating behaviour below room temperature (Sec. \ref{SEC::res}). We attribute this fundamental difference to the presence of much stronger spin-orbit interactions
combined with electronic correlations and crystal-field effects~\cite{aaramkim2017}
that stabilize a spin-orbit Mott insulator with Ir in a 4+ oxidation state. This property
 makes this family a potential platform to explore Kitaev magnetism in a structural framework that we show interpolates between triangular and honeycomb structures.

\begin{figure}
\includegraphics[width=8.5cm]{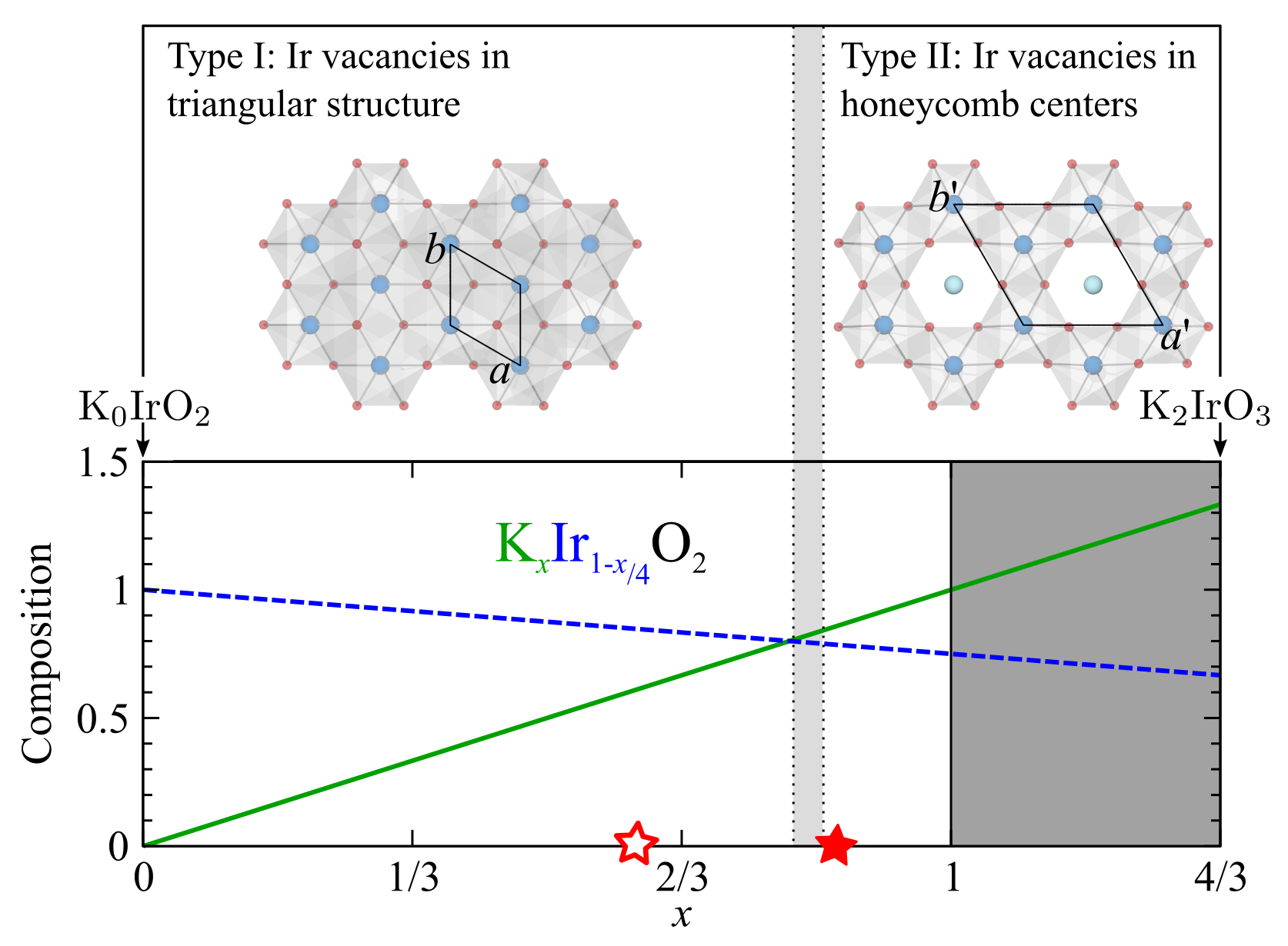}
\caption{\label{FIG::Phase_diagram} Proposed conceptual phase diagram for K$_x$Ir$_{1-x/4}$O$_2$ as a function of the potassium composition $x$. The top diagrams indicate the two structural models for the iridium-oxide layer discussed in the text: Type I, where the Ir vacancies are uniformly distributed on all sites (blue spheres) of a triangular layer, and Type II, where Ir vacancies only occupy the centres (light blue spheres) of a honeycomb structure of filled Ir sites. The vertical grey bar near $x_c=0.82(2)$ indicates the DFT predicted boundary between the two structure types. In the bottom diagram, the blue dashed and solid green lines show the variation of the iridium and potassium compositions, respectively. The solid grey shaded region for $x\in[1,4/3]$ indicates potentially unphysical bonding of potassium in the honeycomb centres. The two representative compositions discussed in detail in the main text are indicated by open and filled red stars on the horizontal bottom axis.}
\end{figure}

We propose the conceptual structural phase diagram shown in
Fig.~\ref{FIG::Phase_diagram}. As noted above, we adopt the hypothesis
that strong spin-orbit coupling combined with electronic correlations and
crystal-field effects favor the 4+ oxidation state for  Ir
atoms. In this case, charge neutrality for different potassium compositions is
maintained by iridium vacancies. Under these constraints we re-write the
chemical formula as K$_x$Ir$_{1-x/4}$O$_2$, such that the phase diagram
(Fig.~\ref{FIG::Phase_diagram}) spans the range of compositions from $x=0$
(K$_0$IrO$_2$) to $x=4/3$ (equivalent to K$_2$IrO$_3$). Consider triangular
layers of edge-sharing IrO$_6$ octahedra, separated by layers of potassium
ions. Such a structure is well established for the related materials
Na$_x$CoO$_2$ \cite{NaxCoO2} and K$_x$CoO$_2$ \cite{Jansen}, where Na/K also
occupy triangular substructures located between successive CoO$_2$ layers. At
$x=0$ we obtain a hypothetical end-member, IrO$_2$. In reality IrO$_2$ adopts a
different crystal structure, in part because a finite amount of potassium is
required to facilitate inter-layer bonding in the proposed layered
K$_x$Ir$_{1-x/4}$O$_2$ structural framework, nonetheless this hypothetical end
member serves as a useful reference. Upon adding potassium and concomitantly
removing some iridium ions (one Ir for every four K added), vacancies are introduced
at random on the nominal Ir sites and the end-member crystal symmetry is
maintained on average. However, at some critical composition (grey vertical
band in Fig.~\ref{FIG::Phase_diagram}) one might expect a structural phase
transition associated with the iridium vacancies forming an ordered
superstructure on the triangular structure. A natural candidate is a tripling
of the unit cell when the vacancies are located preferentially at the centres
of a honeycomb structure. We consider the simplest case of complete preference
of the vacancies to occupy the centre sites, with the surrounding honeycomb
structure being fully occupied. On further increasing the potassium content we
reach a second reference composition at $x=1$ (KIr$_{0.75}$O$_2$), where the K
layer forms a fully-filled triangular layer and the Ir void sites are $1/4$
occupied. Further increasing $x$ beyond $1$ would require placing potassium in
the honeycomb centres. However, this appears to be energetically unfavourable, as it would require a significant distortion of the oxygen environment around the honeycomb centre. The structures with $x<1$ refined below were found to have metal-oxygen bond lengths at the honeycomb centre of about $2.1~\mathrm{\AA}$. These lengths are much shorter than typical K-O bond lengths, for example, in honeycomb K$_2$PbO$_3$ the K-O bond length for K in the honeycomb centres is 2.76 $\mathrm{\AA}$ at room temperature \cite{DELMAS197687}. Despite this, for the purpose of discussion we include in Fig. \ref{FIG::Phase_diagram} compositions with $1 < x \leq 4/3$, with the high end limit
corresponding to the case where all honeycomb centres are filled with K instead
of Ir (nominal composition equivalent to K$_2$IrO$_3$).

In the following, we present a combined single crystal x-ray diffraction (SXD) and density functional theory (DFT) study of the crystal structures adopted by K$_x$Ir$_{1-x/4}$O$_2$. We show that for $0 < x < 1$ two phases exist that form either triangular or honeycomb structures, consistent with the structural framework introduced above. Our DFT calculations show a relatively sharp transition in the region $0.805 < x < 0.843$ between triangular and honeycomb phases, demonstrating that the K:Ir ratio in K$_x$Ir$_{1-x/4}$O$_2$ may act as an effective tuning parameter between the two distinct regimes.

The rest of this paper is organised as follows. In Section \ref{sec::methods} we describe all experimental and theoretical methods used, in the Results sections (\ref{SEC::TypeI} and \ref{SEC::TypeII}) we present the experimental determination of the triangular and honeycomb structures, in the Discussion section (\ref{SEC::Discussion}) we give theoretical support to the above, hypothetical structural phase diagram and demonstrate that the K$_x$Ir$_{1-x/4}$O$_2$ family are candidates to display Kitaev magnetism, and finally we summarise our results in the Conclusions section (\ref{SEC::conclusions}). The appendices include a symmetry analysis of the triangular to honeycomb structural phase transition, $|F_\mathrm{obs}|^2$ plotted against $|F_\mathrm{calc}|^2$ for triangular and honeycomb structural refinements, and structural details on a theoretically proposed K$_2$IrO$_3$ end member compound.

\section{\label{sec::methods} Methods}

Single crystals of K$_x$Ir$_y$O$_2$ (with intended nominal composition $x=2$ and $y=1$) were grown from metallic K and Ir starting materials placed inside Al$_2$O$_3$ crucibles in air, similar to the method used recently for the crystal growth of $\alpha-$Li$_2$IrO$_3$ \cite{Freund}. Ir metal powder was placed at the bottom of the growth crucible, above which Al$_2$O$_3$ fragments were placed to allow nucleation of crystals at sharp edges. K metal was then placed on top of the Al$_2$O$_3$ fragments and the crucible heated to 1070 $^\circ$C in 12 hrs and left there for 70 hrs before cooling to room temperature over 12 hrs.

SXD experiments were performed at room temperature using a laboratory based Oxford Diffraction Supernova diffractometer fitted with a Mo K$_\alpha$ x-ray source. Over thirty crystals (with an approximate diameter of 50-100 $\mu$m) from different synthesized batches were screened to test for consistency in the diffraction patterns. The data sets measured for structural refinement using \textsc{fullprof} \cite{rodriguezcarvaja93} contained approximately full spheres of diffraction peaks in reciprocal space, giving between 2000 and 10000 statistically significant diffraction intensities depending on sample size.

Variable temperature resistance measurements were performed using a Quantum Design Physical Properties Measurement System. A 2-point contact method was employed using $\sim200 \mu\mathrm{m}$ long, parallel silver paint electrodes on the surface of a flat-plate single crystal. The average distance between electrodes was $\sim230 \mu\mathrm{m}$, and a 100$\mu\mathrm{A}$ excitation current was used at all temperatures.

Complementary \emph{ab-initio} density functional theory 
structural relaxation calculations within the proposed structural framework
(Fig.~\ref{FIG::Phase_diagram}), and with variable potassium compositions $x$, were performed 
using the Vienna $Ab-initio$ Simulation Package (VASP)~\cite{Kresse1993, Kresse1996a, Kresse1996b}  adopting 
the Perdew-Burke-Ernzerhof generalized gradient approximation (GGA)~\cite{Perdew1996}  and the
 projector augmented wave method~\cite{Bloechl1994}. In order to study fractional occupancies in K$_x$Ir$_{1-x/4}$O$_2$
 with the possibility of also describing Ir and K vacancies,
 we used the virtual crystal approximation (VCA). 
 In this approach the  crystal structure is kept with the primitive-cell periodicity, but
it is composed of fictitious `virtual' atoms that interpolate between the behavior of the Ir (or K) atoms and vacancies.
  The validity of such an approximation for simulating
 possible vacancy effects has been tested in the past~\cite{MazinVCA2010} and we have
 performed supercell calculations for some computationally feasible cases such as $x=1$.
 After carefully checking the convergence of the results with respect to the cutoff energy and the number of k-points,
 we adopted a cutoff energy of 520 eV and k-points generated by a 12 $\times$ 12 $\times$ 6 
 Monkhorst-Pack grid.
The density of states for theoretically proposed K$_2$IrO$_3$ within GGA including relativistic effects 
were performed with the linearized augmented plane wave (LAPW) method implemented in WIEN2k~\cite{Blaha2001} with a mesh of 1000 ${\bf k}$ points in the first Brillouin zone and RK$_{\rm max}$ was set to 8. The exchange parameters were estimated
using the exact diagonalization method of Ref.~\onlinecite{Winter2016}.

\section{\label{SEC::results}Results}
The SXD diffraction patterns from a range of samples fell into two distinct types: Type I, which could be indexed by a hexagonal unit cell with similar lattice parameters to K$_x$CoO$_2$ \cite{Jansen}, and Type II, which showed additional super-cell peaks attributed to a $\sqrt{3} \times \sqrt{3}$ in-plane supercell. Below we present for each case the key features of the diffraction pattern, the structural refinement analysis, and a detailed comparison with \emph{ab-initio} structural relaxation calculations.\\

\subsection{\label{SEC::TypeI}Type I: Triangular Structure ($x=0.61$)}

\begin{figure*}
\includegraphics[width=18cm]{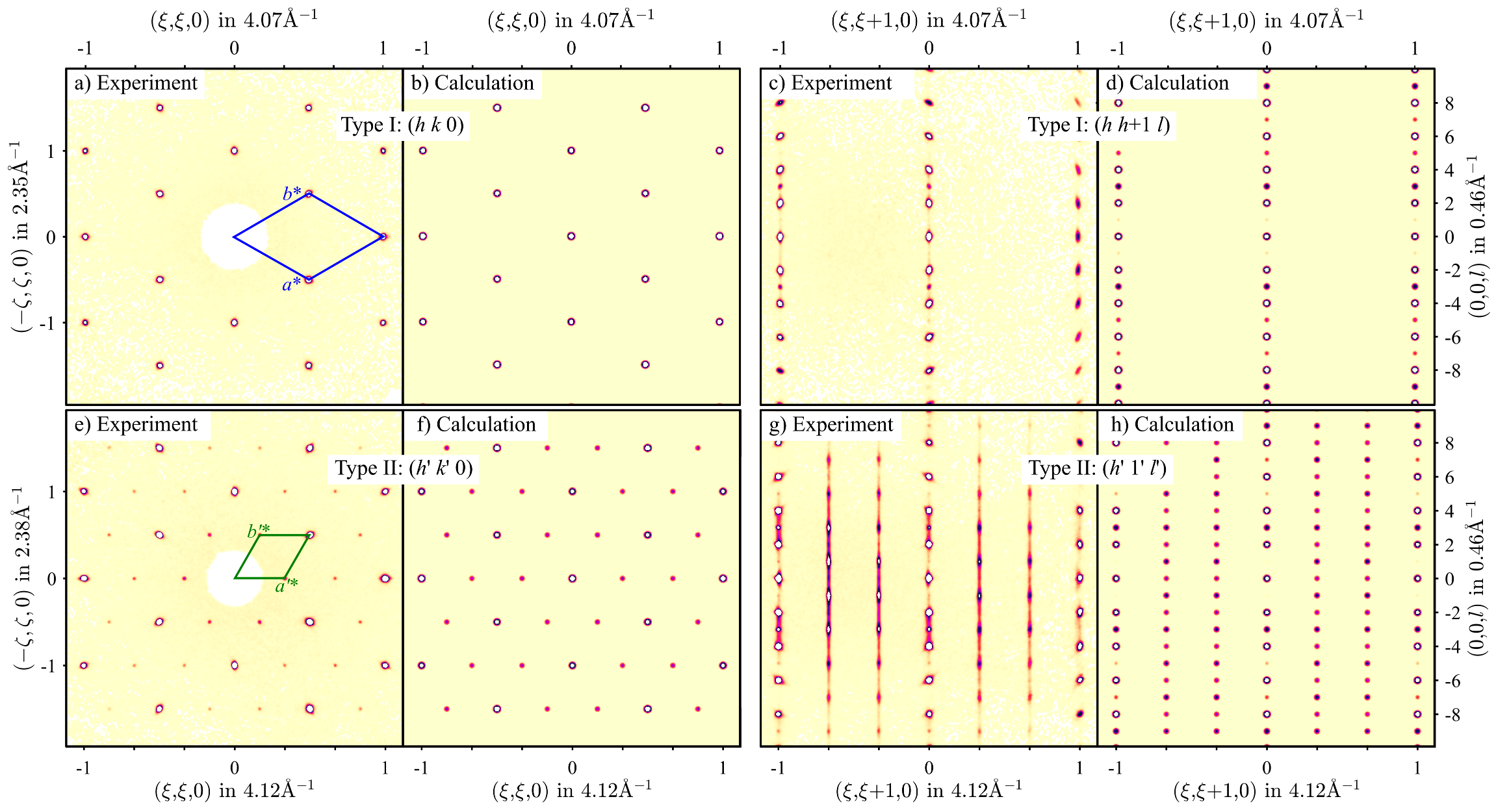}
\caption{\label{FIG::Diffraction_data}Side-by-side experimental vs. calculated x-ray diffraction patterns in two orthogonal reciprocal lattice planes for the Type I (top row) and II (bottom row) crystals described in the text. To facilitate comparison between the two diffraction pattern types the Type II patterns are indexed in the hexagonal ``parent'' cell. The unprimed symbols in the reciprocal plane labels correspond to the Type I structure ($P6_3/mmc$ in Fig.~\ref{FIG::TypeI_structure}) and primed ones to the Type II structure ($P6_322$ in Fig.~\ref{FIG::TypeII_structure}). The reciprocal space unit cells in the two cases are indicated by the blue and green lines in a) and e).}
\end{figure*}

\begin{figure}
\includegraphics[width=8.5cm]{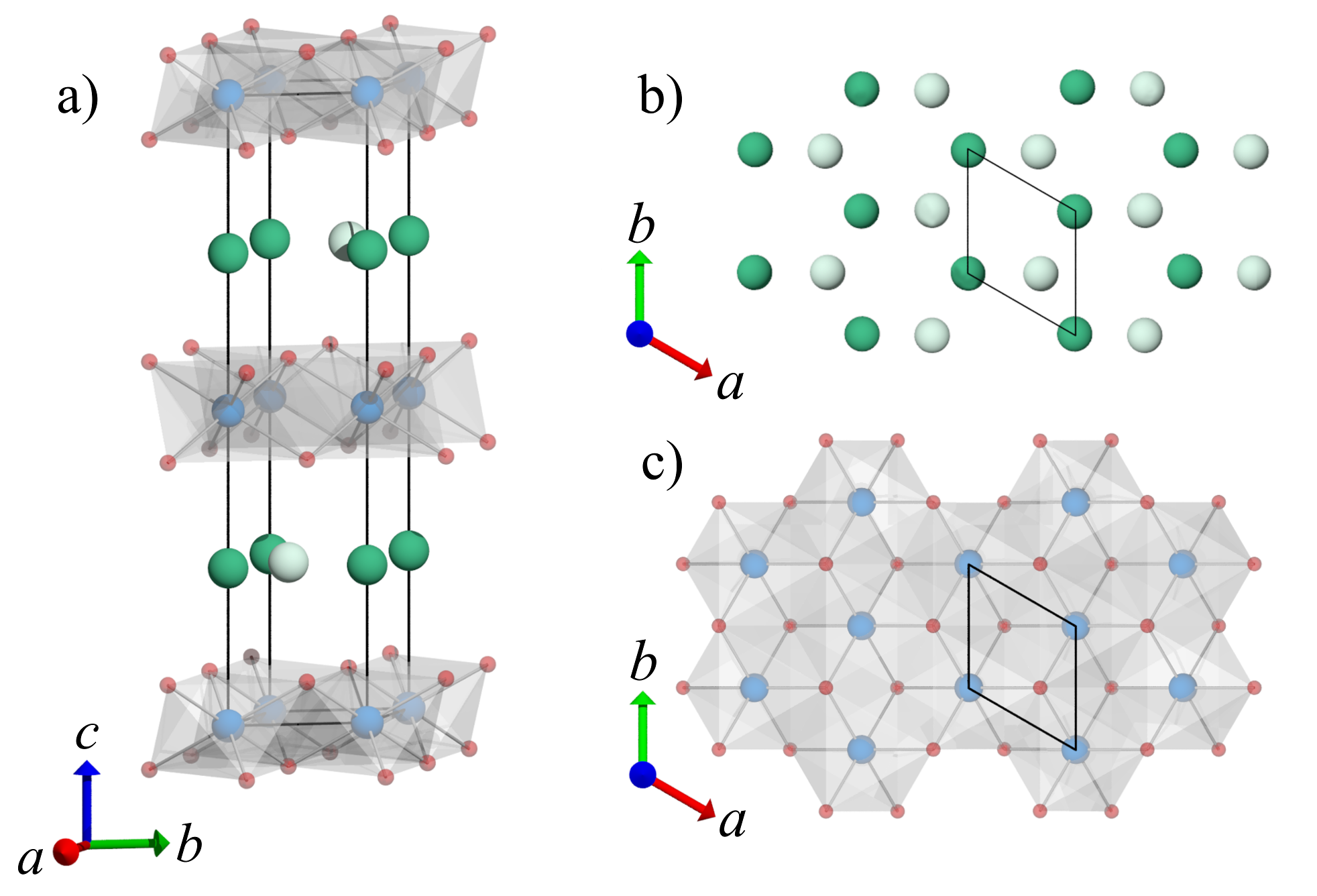}
\caption{\label{FIG::TypeI_structure}The Type I crystal structure of K$_x$Ir$_{1-x/4}$O$_2$ ($x=0.61$) supporting triangular Ir layers. a) A single unit cell drawn with black lines, b) the K1 (dark green spheres) and K2 (light green spheres) potassium triangular substructures at $z=1/4$, and c) the Ir$_{1-x/4}$O$_2$ triangular layer at $z=0$. Iridium and oxygen ions are shown as blue and red spheres, respectively, and IrO$_6$ octahedra are shaded grey.}
\end{figure}

A high quality set of Type I diffraction data measured from a representative single crystal (with a refined composition of $x = 0.61$, see below) (open red star in Fig.~\ref{FIG::Phase_diagram}) was selected for quantitative analysis. Over 99\% of the diffraction peaks could be indexed using a primitive hexagonal unit cell with parameters $a=3.0909(2)~\mathrm{\AA}$ and $c=13.720(2)~\mathrm{\AA}$. Figures \ref{FIG::Diffraction_data}a) and c) show the diffraction patterns in $(h,k,0)$ and $(h,h+1,l)$ reciprocal lattice planes, respectively, where wavevector components are expressed with reference to the above hexagonal unit cell metric. The key features can be naturally explained by having triangular layers of edge-sharing IrO$_6$ octahedra successively stacked with a 180$^{\circ}$ rotation around $c$, which forms the basic two-layer structural framework of K$_x$CoO$_2$ \cite{Jansen}. In this structural model, the main diffraction peaks ($l$=even) have the dominant contribution from the triangular transition metal substructure, and the weak peaks at $l$ odd and $h-k=3n+1$ or $3n+2$, $n$ integer (see Fig.~\ref{FIG::Diffraction_data}c) are due to the 180$^{\circ}$ rotation of the oxygen octahedra between successive layers. In this structural framework potassium ions can be located between layers bonded to the oxygens above and below, either with a trigonal prismatic coordination as in K$_x$CoO$_2$ \cite{Jansen}, for which there are two symmetry inequivalent triangular substructures (labelled K1 and K2, see dark and light green spheres in Figure \ref{FIG::TypeI_structure}b, respectively), or with a linear coordination forming a third possible triangular substructure (labelled K3) similar to O-Ag-O in delafossite $2H$-AgNiO$_2$ \cite{Sorgel}.

To test if the data could be described by the same $P6_3/mmc$ structural model as K$_x$CoO$_2$ data reduction was performed within the respective $6/mmm$ Laue class. This yielded $R_{int}=5.12 \%$, showing this space group to be in excellent quantitative agreement with the diffraction pattern symmetry. The crystal structure model was first tested assuming all sites were fully occupied (test composition KIrO$_2$, with potassium on one of the K1, K2, or K3 triangular substructures). In each case 7 free parameters were refined: the $z$ fractional coordinate of oxygen, 2 anisotropic (in-plane and out-of-plane) atomic displacement parameters (a.d.p.) for each cation, 1 isotropic a.d.p. for oxygen, and a global scale. The case where potassium was linearly coordinated (K3) could be immediately ruled out. Rietveld refinement of the model with potassium located on either K1 or K2 gave similar results with reasonable goodness-of-fit parameters ($R_{F^2}=5.3\%$, $wR_{F^2}=7.6\%$, $R_F=4.2\%$). However, the anisotropic a.d.p.s for potassium always indicated significant delocalisation within the $ab$-plane. 

Occupation factors were then introduced as free parameters for the Ir site and the two potassium substructures (K1 and K2), which led to an excellent fit ($R_{F^2}=3.4\%$, $wR_{F^2}=4.3\%$, $R_F=3.2\%$), and indicated a partial but equal occupation of both K1 and K2 sublattices and a composition consistent with iridium (also partially occupied) being in a 4+ oxidation state. We note that partial occupation of the K1 and K2 sublattices was also found in K$_x$CoO$_2$, suggesting a common feature of this structural framework. A final model was refined against the data, in which the potassium partial occupation $x$ was varied freely but constrained to be the same on both K1 and K2 sublattices, and the iridium occupation constrained to satisfy charge neutrality for valence 4+. The goodness-of-fit ($R_{F^2}=3.5\%$, $wR_{F^2}=4.4\%$, $R_F=3.3\%$) was comparable to the fully unconstrained refinement above, indicating that these constraints adequately describe a minimal model for the Type I crystal structure, as detailed in Table \ref{TAB::triangular_structure} and shown in Figure \ref{FIG::TypeI_structure}. The resulting cation-oxygen bond lengths are also given in Table \ref{TAB::triangular_structure}. The good agreement between the experimental and calculated diffraction pattern is illustrated for the $(h,k,0)$ plane in Fig.~\ref{FIG::Diffraction_data}a) and b), for the $(h,h+1,l)$ plane in Fig.~\ref{FIG::Diffraction_data}c) and d), and for all measured reflections in Figure \ref{FIG::Fcomp_triangular} of Appendix \ref{SEC::Fcomp}.

\begin{table}
\caption{\label{TAB::triangular_structure} Refined RT crystal structure parameters of the K$_{0.61}$Ir$_{0.85}$O$_2$ triangular structure ($R_{F^2}=3.5\%$, $wR_{F^2}=4.4\%$, $R_F=3.3\%$).}
\begin{ruledtabular}
\begin{tabular}{c l c c c c c}
\multicolumn{6}{l}{\textbf{Cell parameters}} \\
\multicolumn{6}{l}{Space group: $P6_3/mmc$ (\#194)} \\
$a,b,c$ ($\mathrm{\AA}$) & 3.0909(2) & 3.0909(2) & 13.720(2) \\
\multicolumn{6}{l}{Volume ($\mathrm{\AA}^3$) 113.52(2)}\\
\\
\multicolumn{6}{l}{\textbf{Atomic fractional coordinates}} \\
Atom & Site & $a$ & $b$ & $c$ & Occ.\\
\hline
Ir & $2a$ & 0 & 0 & 0 & 0.847(8) \\
K1 & $2b$ & 0 & 0 & 1/4 & 0.307(8) \\
K2 & $2d$ & 2/3 & 1/3 & 1/4 & 0.307(8) \\
O  & $4f$ & 1/3 & 2/3 & 0.074(1) & 1 \\
\\
\multicolumn{6}{l}{\textbf{Atomic displacement parameters} ($\mathrm{\AA}^2$)} \\
Ir & \multicolumn{3}{l}{U$_{11}$ = U$_{22}$ = 0.0076(7)} & \multicolumn{2}{l}{U$_{33}$ = 0.0161(7)} \\
K1 & \multicolumn{3}{l}{U$_{11}$ = U$_{22}$ = 0.16(3)} & \multicolumn{2}{l}{U$_{33}$ = 0.03(1)} \\
K2 & \multicolumn{3}{l}{U$_{11}$ = U$_{22}$ = 0.16(3)} & \multicolumn{2}{l}{U$_{33}$ = 0.03(1)} \\
O  & \multicolumn{3}{l}{U$_\mathrm{iso}$ = 0.010(4)}\\
\\
\multicolumn{6}{l}{\textbf{Selected bond lengths} ($\mathrm{\AA}$)} \\
\multicolumn{3}{c}{K1(K2) - O $\quad$ 3.01(1)}\\
\multicolumn{3}{c}{Ir - O $\quad$ 2.050(7)} \\
\\
\multicolumn{6}{l}{\textbf{Data reduction} ($R_\mathrm{int}$: 5.12\%)} \\
\multicolumn{6}{l}{$\#$ measured reflections: 2277} \\
\multicolumn{6}{l}{$\#$ independent reflections $(I > 1.5\sigma$): 88} \\
\multicolumn{6}{l}{$\#$ fitted parameters: 8} \\
\end{tabular}
\end{ruledtabular}
\end{table}

We note that no diffuse scattering is apparent in the diffraction pattern indicating a well-ordered structure with no stacking faults, which are quite common in other layered structures. In the present case the absence of stacking faults may be due to the particular stacking sequence where any in-plane shift would involve a change in the inter-layer bonding geometry.

DFT calculations were employed to test the Type I structural model, which was found to be stable for the refined composition. Relaxation of the atomic fractional coordinates within the experimentally determined unit cell metric gave a small shift in the oxygen $z$ coordinate from the empirical value of 0.074(1) to 0.05896.

To summarise, the Type I crystal structure adopted by the K$_x$Ir$_{1-x/4}$O$_2$ solid solution is composed of triangular layers of edge-sharing Ir$_{1-x/4}$O$_6$ octahedra, where every iridium site is partially occupied and equivalent by the translational symmetry of the lattice. Successive layers are stacked with a 180$^\circ$ rotation around $c$, and potassium ions fill the inter-layer space with an equal, partial occupation of two offset triangular substructures. Finally, the K:Ir composition ratio was found to be consistent with an Ir$^{4+}$ valence.

\subsection{\label{SEC::TypeII}Type II: Honeycomb Structure ($x=0.85$)}

\begin{figure}
\includegraphics[width=8.5cm]{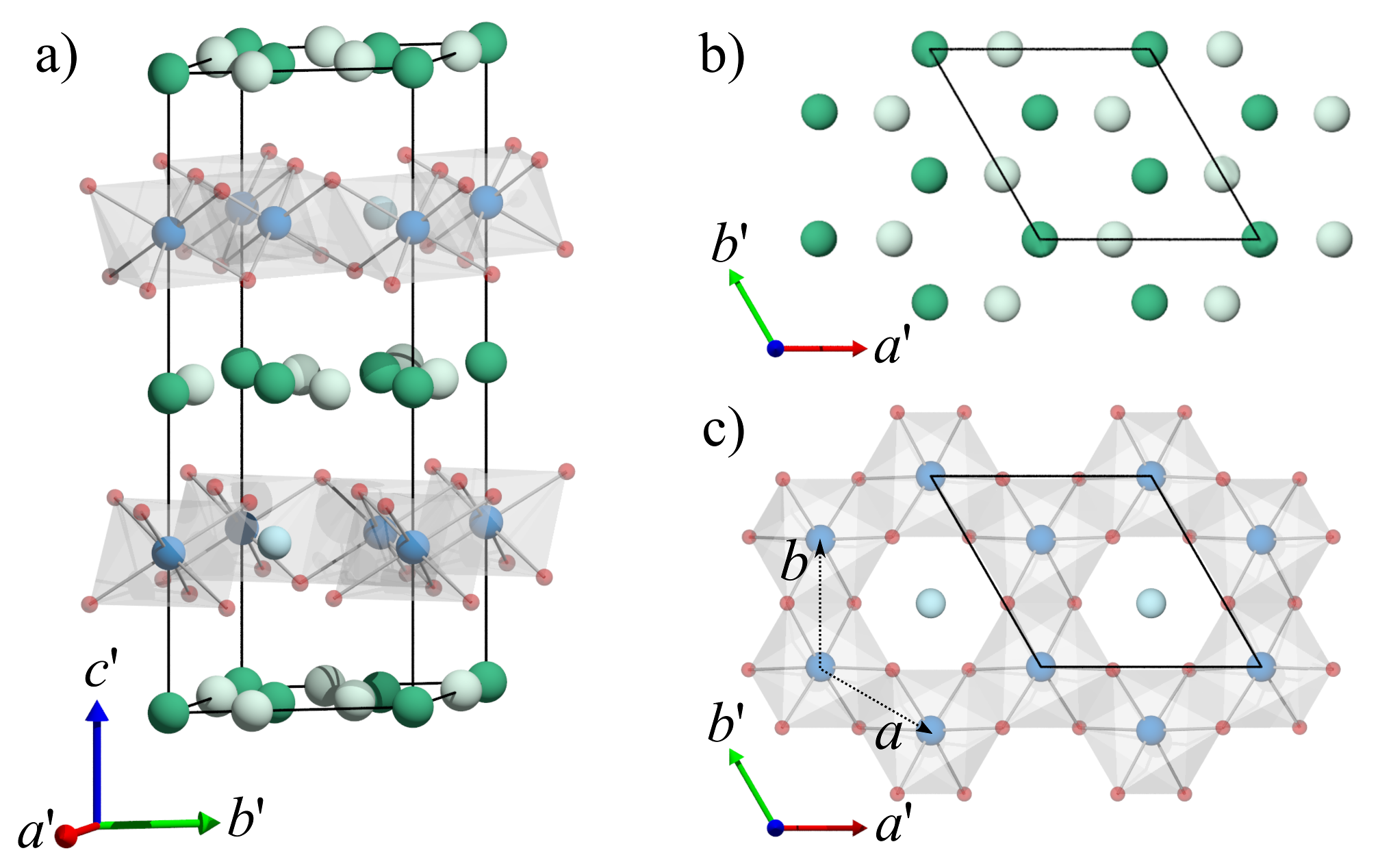}
\caption{\label{FIG::TypeII_structure}The Type II crystal structure of K$_x$Ir$_{1-x/4}$O$_2$ ($x=0.85$) supporting honeycomb Ir layers. a) A single unit cell drawn with black lines, b) the K11, K12 (dark green spheres) and K2 (light green spheres) potassium triangular substructures at $z=0$, and c) the IrO$_2$ triangular layer at $z=1/4$. Iridium, partially occupied iridium, and oxygen ions are shown as blue, light blue, and red spheres, respectively, and fully occupied IrO$_6$ octahedra are shaded grey. For comparison, the lattice basis vectors of the triangular structure are drawn as faint dotted lines.}
\end{figure}

A Type II SXD data set measured from a representative sample
(with a refined composition of $x =0.85$, see below) (filled red star in Fig.~\ref{FIG::Phase_diagram}) was selected for analysis. The diffraction patterns show all the features of Type I, but with additional peaks clearly observed at ($h,k,l$)$\pm$(1/3,1/3,0) positions with $l= 2n+1$, $h$, $k$, $n$ integers, as shown by comparing Fig.~\ref{FIG::Diffraction_data}c) and g). The additional diffraction peaks indicate a tripling of the unit cell in the $ab$ plane. In this case the diffraction data can be indexed with a $\sqrt{3}\times\sqrt{3}\times 1$ super-cell with respect to the ``parent'' Type I hexagonal cell, and refinement against the diffraction data gave the lattice parameters $a'=5.2823(2)~\mathrm{\AA}$ and $c'=13.5437(7)~\mathrm{\AA}$ (primes are used throughout to reference the triple-cell metric). The triple cell diffraction peaks are sizeable in intensity compared to the main triangular structure peaks, indicating that they originate from atomic ordering of iridium ions, which are the dominant scatterer with 77 electrons compared to K (19 electrons) and oxygen (16 electrons). The triple cell can be naturally explained if the iridium vacancies are no longer randomly distributed on all sites of the triangular structure, but are located only at the centres of a honeycomb structure of fully-filled iridium sites, as indicated schematically in Fig.~\ref{FIG::Phase_diagram} top right. The above triple cell metric contains two honeycomb layers, where the honeycomb centres can be stacked either eclipsed (straight on top), or staggered (with an in-plane offset).
The presence of strong triple-cell peaks at odd $l$ indicates a staggered arrangement, as eclipsed stacking would only give triple cell peaks with even $l$. We find that such a structure can be described in the $P6_322$ space group, which is a triple cell super-space group of the triangular structure, $P6_3/mmc$ (in Appendix \ref{SEC::symmetry} we show that this is a unique solution based on symmetry arguments and physical constraints). This structural model is illustrated in Fig.~\ref{FIG::TypeII_structure}. We note that, compared to the $P6_3/mmc$ ``parent'' in Fig.~\ref{FIG::TypeI_structure}, the unit cell is rotated by 30$^{\circ}$ around $c$ and the unit cell origin is shifted by $[0,0,1/4]$. There are now three symmetry-distinct iridium sites, Ir1 (Wyckoff position $2b$), Ir2 ($2c$) and Ir3 ($2d$). Within the $P6_322$ symmetry, Ir1 has eclipsed stacking whereas Ir2 and Ir3 have staggered stacking. Having established that the diffraction pattern is consistent with staggered honeycomb centres, we therefore place the iridium vacancies at the Ir3 ($2d$) site (choosing the Ir2 $2c$ site leads to an equivalent description of the final structure).

The respective $622$ Laue class was used in the data reduction, giving good agreement ($R_\mathrm{int}=6.35\%$). A full structural model was parameterised as given in Table \ref{TAB::honeycomb_structure} and refined against the data. A good agreement with the data was achieved (shown in Figure \ref{FIG::Fcomp_honeycomb} of Appendix \ref{SEC::Fcomp}) with reliability parameters $R_{F^2}=7.0\%$, $wR_{F^2}=9.3\%$, and $R_F=4.4\%$. The resultant structure is summarised in Table \ref{TAB::honeycomb_structure} (including cation-oxygen bond lengths), and is shown in Figure \ref{FIG::TypeII_structure}. As before, the cation occupations were constrained to impose uniform partial occupation of potassium sites (K11 and K12 form one triangular substructure, K2 the other), as well as the 4+ oxidation state of iridium, while allowing the K:Ir ratio to vary. The occupations of Ir sites that compose the honeycomb structure (Ir1 and Ir2) were fixed to one, whilst the site in the honeycomb centre (Ir3) was allowed to adopt a partial occupation. Relaxing these constraints did not significantly improve the fit, showing again that iridium adopts the 4+ oxidation state, which then dictates the iridium occupation of the honeycomb centres for a given amount of potassium in the (K11,K12) and K2 triangular substructures. Comparison  between the experimental and calculated diffraction patterns is illustrated for the $(h',k',0)$ plane in Fig.~\ref{FIG::Diffraction_data}e) and f), and for the $(h',1',l')$ plane in Fig.~\ref{FIG::Diffraction_data}g) and h). All features are well reproduced, with the exception of the diffuse scattering which occurs between successive triple cell peaks along $l$. This diffuse scattering could be naturally understood as arising from faults in the stacking of the honeycomb centres, i.e. if every so often the honeycomb centre would randomly shift from the nominal position to one of the other neighbouring iridium sites. Since this involves no change in the inter-layer bonding geometry this stacking fault would be expected to have a very small energy cost.

\begin{table}
\caption{\label{TAB::honeycomb_structure} Refined RT crystal structure parameters of the K$_{0.85}$Ir$_{0.79}$O$_2$ honeycomb structure ($R_{F^2}=7.0\%$, $wR_{F^2}=9.3\%$, $R_F=4.4\%$).}
\begin{ruledtabular}
\begin{tabular}{l l c c c c c}
\multicolumn{6}{l}{\textbf{Cell parameters}} \\
\multicolumn{6}{l}{Space group: $P6_322$ (\#182)} \\
\multicolumn{6}{l}{$a',b',c'$ ($\mathrm{\AA}$): 5.2823(2), 5.2823(2), 13.5437(7) }\\
\multicolumn{6}{l}{Volume ($\mathrm{\AA}^3$) 327.28(3)}\\
\\
\multicolumn{6}{l}{\textbf{Atomic fractional coordinates}} \\
Atom & Site & $a$ & $b$ & $c$ & Occ.\\
\hline
Ir1 & $2b$ & 0 & 0 & 1/4 & 1 \\
Ir2 & $2c$ & 1/3 & 2/3 & 1/4 & 1 \\
Ir3 & $2d$ & 2/3 & 1/3 & 1/4 & 0.35(8) \\
K11 & $2a$ & 0 & 0 & 0 & 0.43(3) \\
K12 & $4f$ & 1/3 & 2/3 & -0.006(4) & 0.43(3) \\
K2 & $6g$ & 0 & 0.341(9) & 0 & 0.43(3) \\
O  & $12i$ & 0.319(5) & 0.334(5) & 0.1733(9) & 1 \\
\\
\multicolumn{6}{l}{\textbf{Atomic displacement parameters} ($\mathrm{\AA}^2$)} \\
Ir1 & \multicolumn{3}{l}{U$_{11}$ = U$_{22}$ = 0.011(1)} & \multicolumn{2}{l}{U$_{33}$ = 0.022(3)} \\
Ir2 & \multicolumn{3}{l}{U$_{11}$ = U$_{22}$ = 0.0044(6)} & \multicolumn{2}{l}{U$_{33}$ = 0.011(1)} \\
Ir3  & \multicolumn{3}{l}{U$_\mathrm{iso}$ = 0.021(3)} \\
K11, K12 & \multicolumn{3}{l}{U$_{11}$ = U$_{22}$ = 0.13(2)} & \multicolumn{2}{l}{U$_{33}$ = 0.059(9)} \\
K2 & \multicolumn{3}{l}{U$_{11}$ = 0.13(2), U$_{22}$ = 0.08(3)} & \multicolumn{2}{l}{U$_{33}$ = 0.059(9)} \\
O  & \multicolumn{3}{l}{U$_\mathrm{iso}$ = 0.004(3)} \\
\\
\multicolumn{6}{l}{\textbf{Selected bond lengths} ($\mathrm{\AA}$)} \\
\multicolumn{4}{l}{K11 - O $\quad$ 2.91(2)} & \multicolumn{2}{c}{Ir1 - O $\quad$ 2.02(2)} \\
\multicolumn{4}{l}{K12 - O $\quad$ 2.92(5)} & \multicolumn{2}{c}{Ir2 - O $\quad$ 2.01(2)} \\
\multicolumn{4}{l}{K2 - O $\quad~$ 2.90(3), 2.93(3), 2.97(3)} & \multicolumn{2}{c}{Ir3 - O $\quad$ 2.11(3)} \\
\\
\multicolumn{6}{l}{\textbf{Data reduction} ($R_\mathrm{int}$: 6.35\%)} \\
\multicolumn{6}{l}{$\#$ measured reflections: 9755} \\
\multicolumn{6}{l}{$\#$ independent reflections $(I > 1.5\sigma$): 234} \\
\multicolumn{6}{l}{$\#$ fitted parameters: 8} \\
\end{tabular}
\end{ruledtabular}
\end{table}

The composition of this sample was refined to K$_{0.85}$Ir$_{0.79}$O$_2$. We note that, compared to the Type I sample reported above, the increased potassium content is consistent with the contraction of the $c$ lattice parameter through greater inter-layer bonding (the same effect is observed in Na$_x$CoO$_2$ \cite{NaxCoO2}). Furthermore, the $ab$ plane was found to contract upon formation of the honeycomb layers. DFT calculations were used to test the $P6_322$ structural model, which was found to be stable for the refined composition. Relaxation of the atomic fractional coordinates within the experimentally determined unit cell metric gave small shifts ($<3\sigma$) in all but one of the freely varying atomic fractional coordinates. A significant shift in the K12 $z$ coordinate, located directly above or below the honeycomb centre, was found to be strongly dependent upon the iridium occupation of the honeycomb centre (Ir3) site. A DFT relaxation performed on model structures with extremal compositions K$_{0.1}$Ir$_{0.975}$O$_2$ (0.925 Ir3 occupation) and K$_{1}$Ir$_{0.75}$O$_2$ (0.25 Ir3 occupation) gave K12 $z$ coordinates of -0.015 and -0.058, respectively, clearly indicating that in theory the K12 potassium ion relaxes towards the honeycomb centre in the absence of an Ir3 ion. It was therefore surprising to find that the experimental K12 $z$ coordinate was refined to -0.006(4), despite the relatively low occupation of the Ir3 site. This discrepancy can be reconciled by considering the preferential occupation of (K11,K12) or K2 triangular substructures for both extremes of Ir3 occupation. Further DFT relaxations demonstrated that the K2 site is preferentially occupied when the Ir3 occupation is low, and that the (K11,K12) sites are preferentially occupied when the Ir3 occupation is close to one. Although we have to consider an average occupation of potassium sites in the SXD data analysis, it is unphysical for both to be (partially) occupied in the same unit cell. A more realistic picture is one in which extended regions of the sample have potassium located in \emph{either} the (K11,K12) \emph{or} the K2 sites. If the former region also has an Ir3 occupation close to one, as indicated by DFT, the K12 ion never has the opportunity to relax into the honeycomb centres and its $z$ coordinate will refine to $\sim0$ on average, independent of the average Ir3 site occupation.

In summary, the Type II crystal structure adopted by the K$_x$Ir$_{1-x/4}$O$_2$ solid solution is composed of honeycomb layers of edge-sharing Ir$^{4+}$O$_6$ octahedra, where every iridium site of the honeycomb is fully occupied, but the honeycomb centre has a reduced iridium occupation. Neighbouring honeycomb layers are related by a 180$^{\circ}$ rotation. Potassium ions fill the inter-layer space with an equal, partial occupation of two triangular substructures such that for a given K:Ir compositional ratio iridium maintains a 4+ oxidation state.

We note that the Type II structure is different to that of other structures of stacked honeycombs of edge-shared IrO$_6$ octahedra. In particular, in the layered iridates Na$_2$IrO$_3$ \cite{Choi12,Chun15} and $\alpha$-Li$_2$IrO$_3$ \cite{Williams} successive honeycombs are stacked with a \emph{constant} in-plane offset \emph{perpendicular} to an Ir-Ir bond, leading to an overall monoclinic (C2/m) structure, with a typical spacing between adjacent layer centres of $5.3~\mathrm{\AA}$ and $4.8~\mathrm{\AA}$, respectively. To the contrary, in honeycomb K$_x$Ir$_y$O$_2$ successive honeycomb centres are stacked with an \emph{alternating} in-plane offset \emph{parallel} to an Ir-Ir bond, resulting in a two-layer hexagonal structure ($P6_322$, Fig. \ref{FIG::TypeII_structure}), with a spacing between layer centres of $6.8~\mathrm{\AA}$.

\begin{figure}
\includegraphics[width=8.5cm]{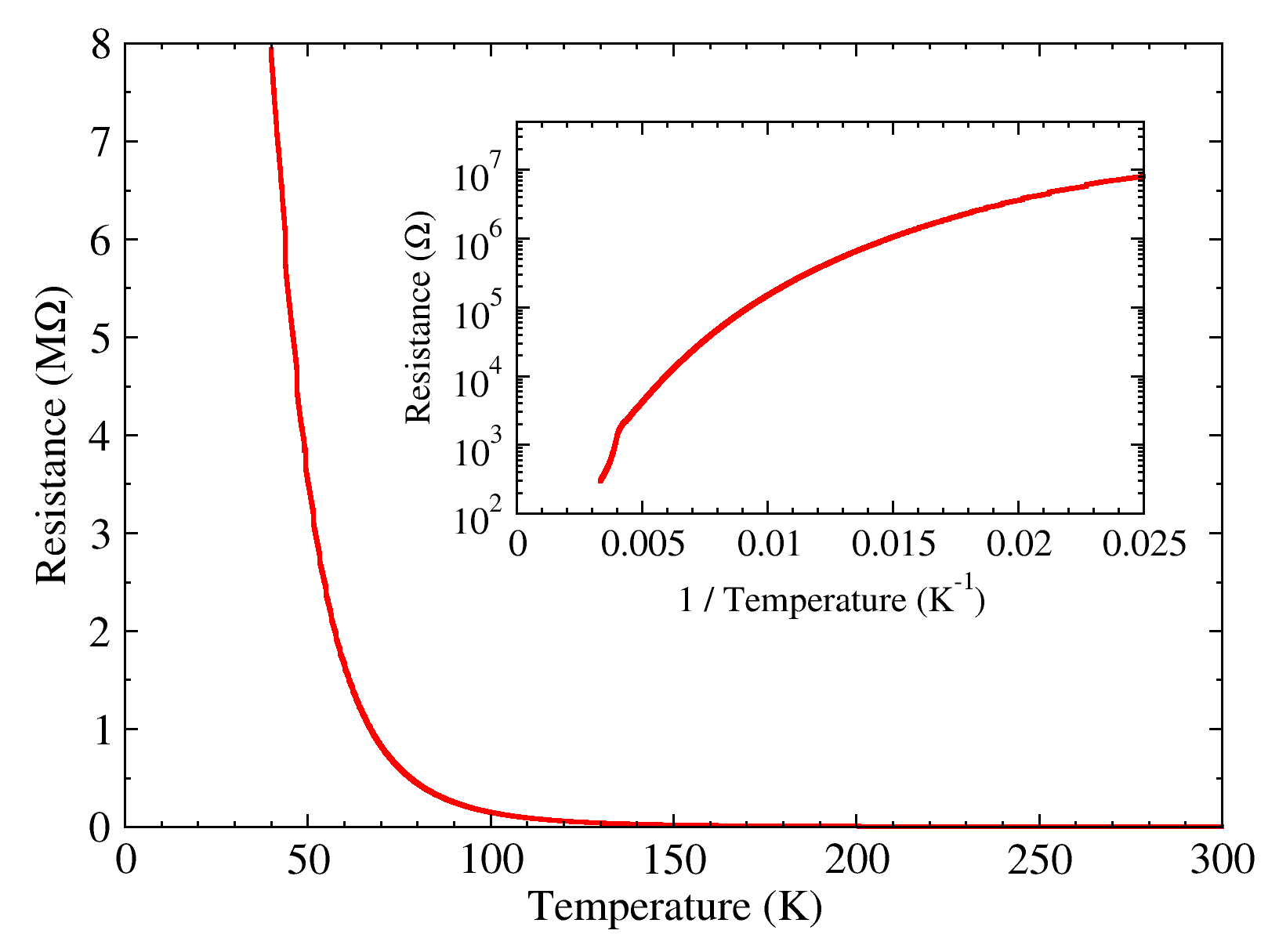}
\caption{\label{FIG::res}The resistance of K$_x$Ir$_{1-x/4}$O$_2$ measured as a function of temperature at the surface of a single crystal sample ($\sim230 \mu\mathrm{m}$ separation between $\sim200 \mu\mathrm{m}$ long parallel contacts). The inset shows the resistance plotted on a logarithmic scale against inverse temperature.}
\end{figure}

\subsection{\label{SEC::res}Resistivity}

The DC electrical resistance at the surface of a K$_x$Ir$_{1-x/4}$O$_2$ single crystal was measured as a function of temperature (Figure \ref{FIG::res}). The resistance was found to rapidly increase upon cooling, clearly demonstrating insulating behaviour, and hence directly supporting our previous hypothesis that the K$_x$Ir$_{1-x/4}$O$_2$ materials are Mott insulators. The inset to Figure \ref{FIG::res} shows the same resistance data on an Arrhenius plot. The non-linearity indicates a departure from 
conventional activated behaviour, as was observed for the related spin-orbit Mott insulator Na$_2$IrO$_3$ \cite{singh2010}. An anomalous drop in the resistance above $\sim255$K is visible in the Figure \ref{FIG::res} inset near 1/T = 0.004 K$^{-1}$. No such anomaly was observed in Na$_2$IrO$_3$ \cite{singh2010}, which may indicate a cross-over between two different conduction processes, and we note that the partial occupation of potassium sites may introduce significant ionic conductivity in K$_x$Ir$_{1-x/4}$O$_2$.

\begin{figure}
\includegraphics[width=8.5cm]{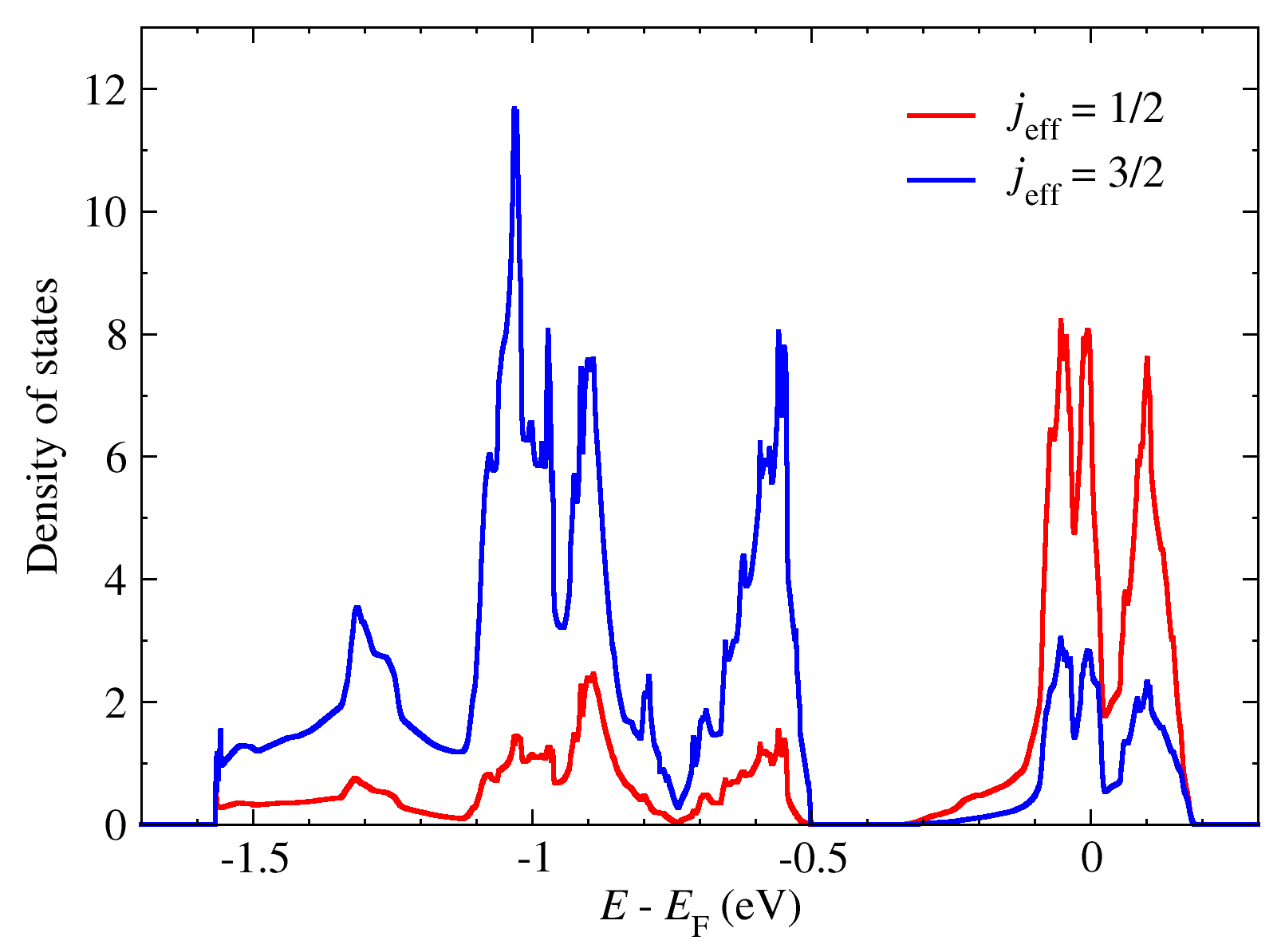}
\caption{ Density of states (DOS) for theoretical K$_2$IrO$_3$ obtained within GGA+SO projected onto the
$j_{\rm eff}$ basis.}
\label{FIG:dos}
\end{figure}

\section{\label{SEC::Discussion}Discussion}

In order to further substantiate 
our proposed phase diagram (Fig.~\ref{FIG::Phase_diagram}) we performed a series of DFT
calculations to explore the boundary between Type I and Type II structures, as well as the structures of
the end systems (IrO$_2$ and
the hypothetical K$_2$IrO$_3$).
For the former case, we carried out structural relaxations of K$_x$Ir$_{1-x/4}$O$_2$ at various fractional occupancies between
x = 0.616 and  0.868 within the VCA approximation as described in Section \ref{sec::methods}. While for $x < 0.805$ the
calculations converge into a Type I
structure, for $ x > 0.843$  the relaxed structures are compatible with a Type II structure establishing the
region between $0.805 < x < 0.843$ as the phase boundary between the two structural models. Interestingly,
 the transition region $x_c$ is close to equal amounts of Ir and K.

The analysis of the end compounds in Fig.~\ref{FIG::Phase_diagram} brings a few new aspects into our study. For IrO$_2$
we found that the most frequently adopted rutile structure~\cite{goldschmidt1926} is only about 224.4 meV/atom (in GGA+SOC
calculations) more stable than the (hypothetical) Type I structure consisting of triangular layers of Ir. Actually,
similar conclusions were drawn by a recent study on a two-dimensional variant of IrO$_2$~\cite{Boeri2019}.
Our observation suggests that a structural phase transition between the Type I structure and
the rutile structure is to be expected at very low $x$ values.
On the other hand, for the far right hypothetical end compound K$_2$IrO$_3$, 
DFT structural relaxation calculations suggest that the honeycomb structure is locally stable (for the predicted structural parameters see Table \ref{TAB::k2ir03_structure} in Appendix \ref{SEC::k2ir03_structure}) even though relatively small K-O bond lengths of 2.45 \AA~ at the honeycomb centre are found (for comparison, the K-O bond length at the honeycomb centres in K$_2$PbO$_3$ is 2.76 $\mathrm{\AA}$ at room temperature \cite{DELMAS197687}).
 While this structure may not be realizable at ambient conditions,
 it could be explored by high-pressure synthesis techniques. The electronic properties of the theoretically-proposed K$_2$IrO$_3$
 reveal a similar spin-orbit-assisted-Mott-insulator behavior to the  layered honeycomb  materials
 $\alpha$-Li$_2$IrO$_3$ \cite{winter2016challenges}, Na$_2$IrO$_3$ \cite{foyevtsova2013ab} and $\alpha$-RuCl$_3$ \cite{johnson2015monoclinic}.  Fig.~\ref{FIG:dos} shows the relativistic (GGA+SO) density of states to be dominantly
$j_{\rm eff}=1/2$ with a small contribution from $j_{\rm eff}=3/2$ at the Fermi level.  Addition of a small onsite
Coulomb repulsion U (not shown)
opens accordingly a gap in the density of states.

Based on the relativistic basis, we calculated the effective $S = j_{\rm eff}=1/2$  model for the Z-bond
\begin{align}
H_{\rm ij} = J \mathbf{S}_i \cdot \mathbf{S}_j + KS^zS^z + \Gamma (S^x S^y + S^y S^x)\\
 + \Gamma^{\prime} (S^x S^z + S^y S^z + S^z S^x + S^z S^y),
\end{align}
and found ($J$, $K$, $\Gamma$, $\Gamma^{\prime}$) $\sim$ (-3.2, -23.2, -1.6, 9.7) meV. Remarkably, these values situate this system
near the Kitaev limit. 

Finally, as discussed in Section \ref{SEC::results}, our relativistic DFT calculations for K$_x$Ir$_{1-x/4}$O$_2$ for all $x$ values
are compatible with the assumption of a Ir$^{4+}$ oxidation state and the presence of vacancies. In this case,
previous model calculations~\cite{aaramkim2017} including spin-orbit coupling, electronic correlations and
crystal-field effects show that these systems are in the regime of spin-orbit Mott insulators.

\section{\label{SEC::conclusions}Conclusion}
In summary, we have reported structural studies of recently-synthesized single crystals of K$_x$Ir$_y$O$_2$, with potassium intercalating triangular layers of edge-sharing IrO$_6$ octahedra. We have proposed that the strong spin-orbit coupling of iridium cations disfavours fractional Ir valence and metallic conductivity (as observed in the iso-structural families with Ir replaced by Co or Ru), but instead stabilises a spin-orbit Mott insulator with all Ir sites in the 4+ oxidation state. In this case, charge neutrality for different potassium compositions, $x$, is achieved via the introduction of Ir vacancies, $y=1-x/4$. We have presented structural relaxation calculations that predict that the vacancies are randomly distributed over the triangular Ir substructure up to a critical concentration, $x_c$, above which they order at the centres of a honeycomb structure of fully-occupied Ir sites. The two structural models either side of $x_c$ were refined against representative x-ray diffraction data sets, which were both found to be in excellent quantitative agreement with ab-initio calculations. Our results suggest that K$_x$Ir$_{1-x/4}$O$_2$ may realize a platform to explore Kitaev magnetism of strong spin-orbit coupled magnetic moments in a structural framework that interpolates between triangular and honeycomb structures.

\begin{acknowledgments}
RDJ acknowledges support from a Royal Society University Research Fellowship, RV acknowledges support from the Deutsche Forschungsgemeinschaft (DFG) through grant VA117/15-1, IB acknowledges support from a University of Oxford Clarendon Fund Scholarship, and KM acknowledges University Grants Commission - Council of Scientific \& Industrial Research India for a fellowship. This research was partially supported by the European Research Council (ERC) under the European Union’s Horizon 2020 research and innovation programme Grant Agreement Number 788814 (EQFT) and by the EPSRC (UK) under Grant No. EP/M020517/1. 
\end{acknowledgments}

\bibliography{KIO_bibliography}

\appendix
\section{\label{SEC::symmetry}Symmetry Analysis}

In this Appendix we present a symmetry analysis of the phase transition from the parent triangular structure (Type I) to the honeycomb super-structure (Type II), based on symmetry-adapted modes of atomic site occupations, and we conclude that the proposed $P6_322$ space group is the unique solution to describe the Type II structure. The analysis calls upon two results drawn directly from the diffraction data presented in the main text. Firstly, the triple cell peaks can be indexed using the propagation vector $\mathbf{k}=\pm(1/3,1/3,0)$, defined with respect to the parent hexagonal lattice. Secondly, the intensity of the triple cell peaks is of a similar order of magnitude to the main diffraction peaks, indicating that their origin lies in a modulation of iridium site occupancies. 

We define the occupation of iridium site $j$ as
\begin{equation}
O_j = \sum_n A_n \nu_{n,j}
\end{equation}
where $A_n$ is the amplitude of mode $n$, $\nu_{n,j}$ is the scalar value of mode $n$ at site $j$, and the sum is taken over all modes. There are 6 iridium sites in the $\sqrt{3}\times\sqrt{3}\times 1$ super-cell, for which there are 4 physically real, linearly independent symmetry-adapted modes for $\mathbf{k}=(1/3,1/3,0)$, and 2 linearly independent symmetry-adapted modes for $\mathbf{k}=(0,0,0)$, as calculated using \textsc{isodistort}\cite{Campbell06, Stokes07} and given in Table \ref{TAB::modes}.

\begin{table*}
\caption{\label{TAB::modes} Symmetry-adapted modes of the occupation of the six iridium sites in the $\sqrt{3}\times\sqrt{3}\times 1$ super-cell.}
\begin{ruledtabular}
\begin{tabular}{c | c c | c c c c | c c}
& \multicolumn{2}{c|}{Fractional Coordinates} & \multicolumn{4}{c|}{$\mathbf{k}=(1/3,1/3,0)$ modes} & \multicolumn{2}{c}{$\mathbf{k}=(0,0,0)$ modes} \\
Site & Parent & $P6_322$ super-cell & $\nu_1$ & $\nu_2$ & $\nu_3$ & $\nu_4$ & $\nu_5$ & $\nu_6$ \\
\hline
Ir1\_1 & $0,0,0$   & $0,0,1/4$       & 1    & 0    & 1    & 0    & 1 & 1  \\
Ir2\_1 & $0,1,0$   & $1/3,2/3,1/4$   & -1/2 & -1/2 & -1/2 & -1/2 & 1 & 1  \\
Ir3\_1 & $1,1,0$   & $2/3,1/3,1/4$   & -1/2 & 1/2  & -1/2 & 1/2  & 1 & 1  \\
Ir1\_2 & $0,0,1/2$ & $0,0,3/4$     & 1    & 0    & -1   & 0    & 1 & -1 \\
Ir2\_2 & $1,1,1/2$ & $2/3,1/3,3/4$ & -1/2 & 1/2  & 1/2  & -1/2 & 1 & -1 \\
Ir3\_2 & $0,1,1/2$ & $1/3,2/3,3/4$ & -1/2 & -1/2 & 1/2  & 1/2  & 1 & -1     
\end{tabular}
\end{ruledtabular}
\end{table*}

We can assume that the phase transition between triangular and honeycomb structures is driven by a primary order parameter (OP) associated with an irreducible representation (irrep) of $\mathbf{k}=(1/3,1/3,0)$. This primary OP may couple to secondary, structural OPs associated with other $\mathbf{k}=(1/3,1/3,0)$ or $\mathbf{k}=(0,0,0)$ irreps (\emph{i.e.} coupling to the totally symmetric representation, $\Gamma_1^+$, is always allowed by symmetry). The $K$-point $(1/3,1/3,0)$ reducible representation for the atomic occupation of the iridium sites (Wyckoff site 2a in the parent space group $P6_3/mmc$), decomposes into two 2D irreducible representations, $K_1$ and $K_2$. Similarly, the $\Gamma$-point $(0,0,0)$ reducible representation decomposes into two 1D irreducible representations, $\Gamma_1^+$ and $\Gamma_3^+$. 

Using \textsc{isodistort}\cite{Campbell06, Stokes07} we have determined all super-space groups of the parent $P6_3/mmc$ structure that correspond to the symmetry-distinct directions of the primary OPs in the space spanned by the $K_1$ and $K_2$ irreps (Table \ref{TAB::superspacegroups}). In all cases the $\sqrt{3}\times\sqrt{3}\times 1$ super-cell is defined using the basis $(2,1,0),(-1,1,0),(0,0,1)$ with respect to the parent. For all super-space groups there is no origin shift, except for $P6_322$ where the origin is shifted by (0,0,1/4) with respect to the parent. For each super-space group listed in Table \ref{TAB::superspacegroups} we give the irrep. of the primary symmetry breaking OP and the respective symmetry-adapted mode(s). We also list the secondary modes (and their irreps., given in parenthesis), which are symmetry allowed within the given super-space group and can couple to the primary order parameter. We note that if the phase transition involves secondary modes it will necessarily be of first order. 

\begin{table}
\caption{\label{TAB::superspacegroups}Super-space groups of the parent $P6_3/mmc$ structure that correspond to the symmetry-distinct directions of the primary OPs in the space spanned by the $K_1$ and $K_2$ irreps, and the active symmetry-adapted modes. The irreducbile representations of the symmetry allowed secondary modes are given in parenthesis.}
\begin{ruledtabular}
\begin{tabular}{l | l l l}
Space & Irreducible           & Primary & Secondary \\
group & representations       & modes   & modes \\
\hline
$P6_3/mcm$   & $K_1$ ($\Gamma_1^+$)        & $\nu_1$          & $\nu_5$  \\
$P\bar{6}c2$ & $K_1$ ($\Gamma_1^+$)        & $\nu_2$          & $\nu_1$, $\nu_5$  \\
$P6_322$     & $K_2$ ($K_1$, $\Gamma_1^+$) & $\nu_4$          & $\nu_1$, $\nu_5$  \\
$P\bar{3}1m$ & $K_2$ ($K_1$, $\Gamma_1^+$, $\Gamma_3^+$) & $\nu_3$  & $\nu_1$, $\nu_5$, $\nu_6$  \\ 
$P312$       & $K_2$ ($K_1$, $\Gamma_1^+$, $\Gamma_3^+$) & $\nu_3$, $\nu_4$ & $\nu_1$, $\nu_2$, $\nu_5$, $\nu_6$ 
\end{tabular}
\end{ruledtabular}
\end{table}

As shown in the main text, the Type II honeycomb structure can be identified by additional diffraction peaks when compared to the diffraction pattern of the parent, Type I structure. These additional peaks satisfy the reflection conditions $h' - k' = 3n+1$ or $3n+2$ and $l' = 2m+1$ ($n$,$m$ integers) where, as in the main text, primed indices indicate that the wavevector components are defined with respect to the $\sqrt{3}\times\sqrt{3}\times 1$ super-cell. Corresponding diffraction reflection conditions derived from structure factor calculations of the symmetry adapted modes are given in Table \ref{TAB::conditions}. We find that the observed reflection conditions are only consistent with either $\nu_3$ or $\nu_4$ being the primary OP mode. We can rule out $\nu_3$ as it corresponds to an unphysical imbalance in the total iridiuim occupaction of neighbouring layers, and hence we find this symmetry analysis to be in full support of the experimentally determined Type II honeycomb structure with space group $P6_322$ presented in the main text.

\begin{table}
\caption{\label{TAB::conditions}Diffraction reflection conditions for the symmetry-adapted modes, defined with respect to the reciprocal lattice of the $\sqrt{3}\times\sqrt{3}\times 1$ super-cell.}
\begin{ruledtabular}
\begin{tabular}{l | l l}
Mode & \multicolumn{2}{c}{Reflection conditions ($n$,$m$ integers)} \\
\hline
$\nu_1$ & $h' - k' = 3n+1$ or $3n+2$ & $l' = 2m$   \\
$\nu_2$ & $h' - k' = 3n+1$ or $3n+2$ & $l' = 2m$   \\
$\nu_3$ & $h' - k' = 3n+1$ or $3n+2$ & $l' = 2m+1$ \\
$\nu_4$ & $h' - k' = 3n+1$ or $3n+2$ & $l' = 2m+1$ \\
$\nu_5$ & $h' - k' = 3n$ & $l' = 2m$  \\
$\nu_6$ & $h' - k' = 3n$ & $l' = 2m+1$    
\end{tabular}
\end{ruledtabular}
\end{table}

\begin{figure}
\includegraphics[width=8.0cm]{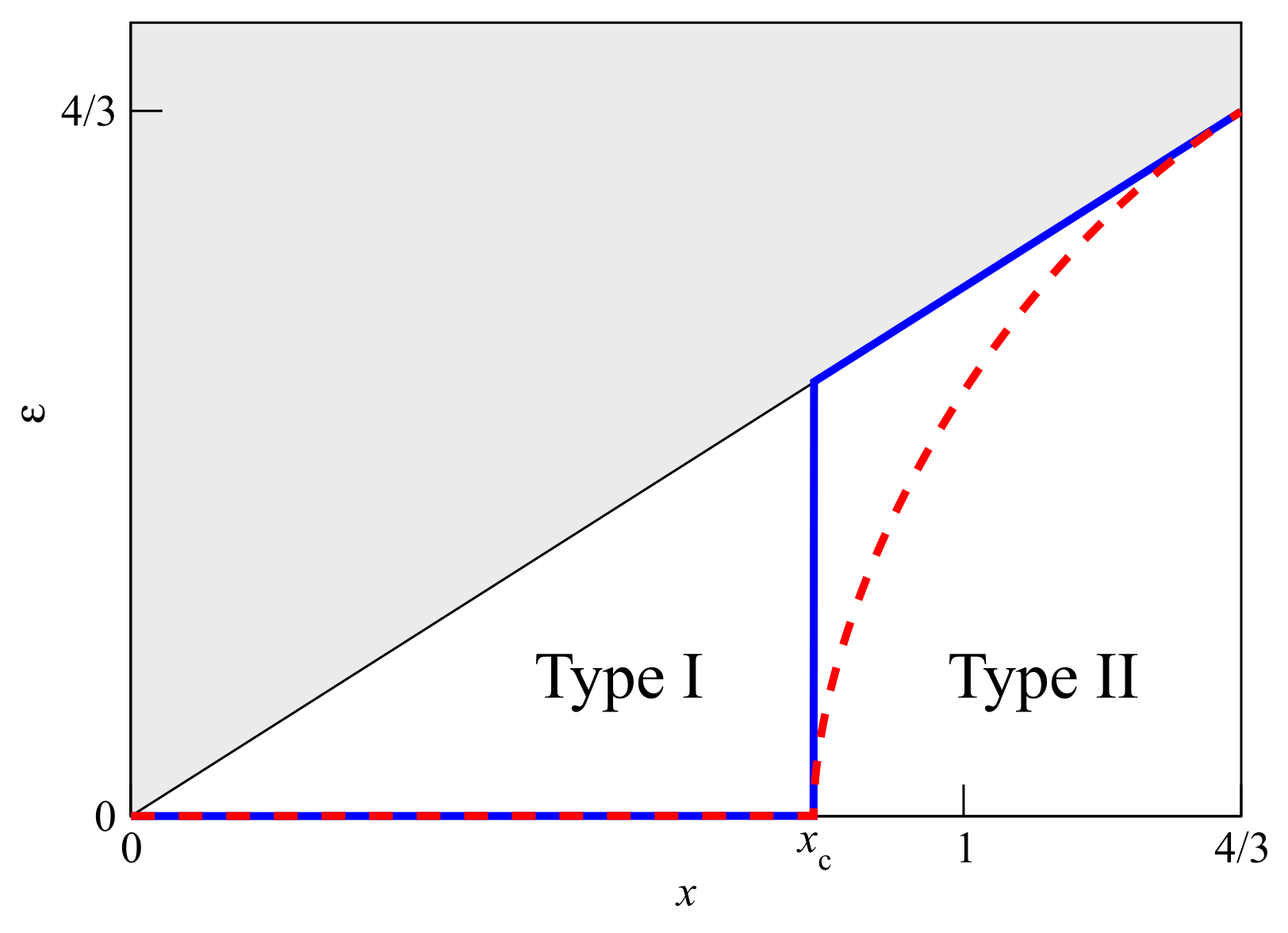}
\caption{\label{FIG::sym_pd}Schematic representation of the order parameter, $\epsilon$, for iridium vacancy ordering, as a function of potassium concentration, $x$, for two generic cases (blue solid and red dashed lines). Below the phase transition at $x_\mathrm{c}$, iridium vacancies are uniformly distrbuted on the triangular structure ($\epsilon=0$). Above $x_\mathrm{c}$, $\epsilon$ may jump to its maximum allowed value for a given $x$ (solid blue line, $\epsilon = x$), corresponding to a fully occupied honeycomb structure with partially occupied honeycomb centres. The red dashed line illustrates another possible scenario where $\epsilon$ increases gradually corresponding to a structure with a finite but not complete preferential vacancy occupation at honeycomb centres. The solid black line passing through the origin ($\epsilon=x$) marks the limit above which (grey shaded region) iridium site occupations are unphysical ($>1$). }
\end{figure}

We can now construct a symmetry-based model for the occupation of the iridium sites that fully captures the structural evolution of K$_x$Ir$_{1-x/4}$O$_2$ as a function of $x$, from the parent Type I $P6_3/mmc$ phase to the lower symmetry, Type II $P6_322$ phase, as hypothesised in the Introduction. Consider the linear combination of $\nu_1$, $\nu_4$, and $\nu_5$ modes,
\begin{equation}\label{EQN::modes_model}
O_j = \left(1-\frac{x}{4}\right)\nu_{5,j} + \frac{\epsilon}{4}(\nu_{1,j} \pm 3\nu_{4,j})
\end{equation}
where the parameter $x$ is the potassium composition. The amplitude $(1-x/4)$  describes the $average$ iridium site occupation, and $\epsilon$ is the order parameter describing the Type I to Type II structural phase transition (the $\pm$ leads to two equivalent descriptions of the Type II structure related by an origin shift). The prefactor of the second term is chosen such that $\epsilon \leq x$ (see below). In the Type I phase $\epsilon = 0$, and the iridium sublattice is uniformly depleated. Just above a critical composition, $x_\mathrm{c}$, the parameter $\epsilon \neq 0$ and a phase transition to the Type II structure occurs. This parameter space is summarised in Figure \ref{FIG::sym_pd}. It is required that $\epsilon \leq x$ to avoid unphysical site occupancies $>1$ (grey shaded region Figure \ref{FIG::sym_pd}). Immediately above $x_\mathrm{c}$, $\epsilon$ may jump to its maximium value and continue to grow as $\epsilon = x$. In this case the honeycomb structure is always fully occupied, with all iridium vacancies located at the honeycomb centres (blue solid line in Figure \ref{FIG::sym_pd}). Alternatively, $\epsilon$ may grow gradually (red dashed line in Figure \ref{FIG::sym_pd}), giving a uniformly depleated honeycomb structure, but with vacancies preferentially located at the honeycomb centres. In both cases the phase transition is first order, but weakly so in the latter. We note that the x-ray diffraction data cannot distinguish between the two scenarios, so for concreteness in the analysis presented in the main text we assumed that iridium vacancies are located only at the honeycomb centres (blue line scenario in Fig. 6) at site Ir3, which corresponds to the minus sign in the second term of Equation \ref{EQN::modes_model}.

\section{\label{SEC::Fcomp}X-ray diffraction goodness-of-fit}

\begin{figure}
\includegraphics[width=8.5cm]{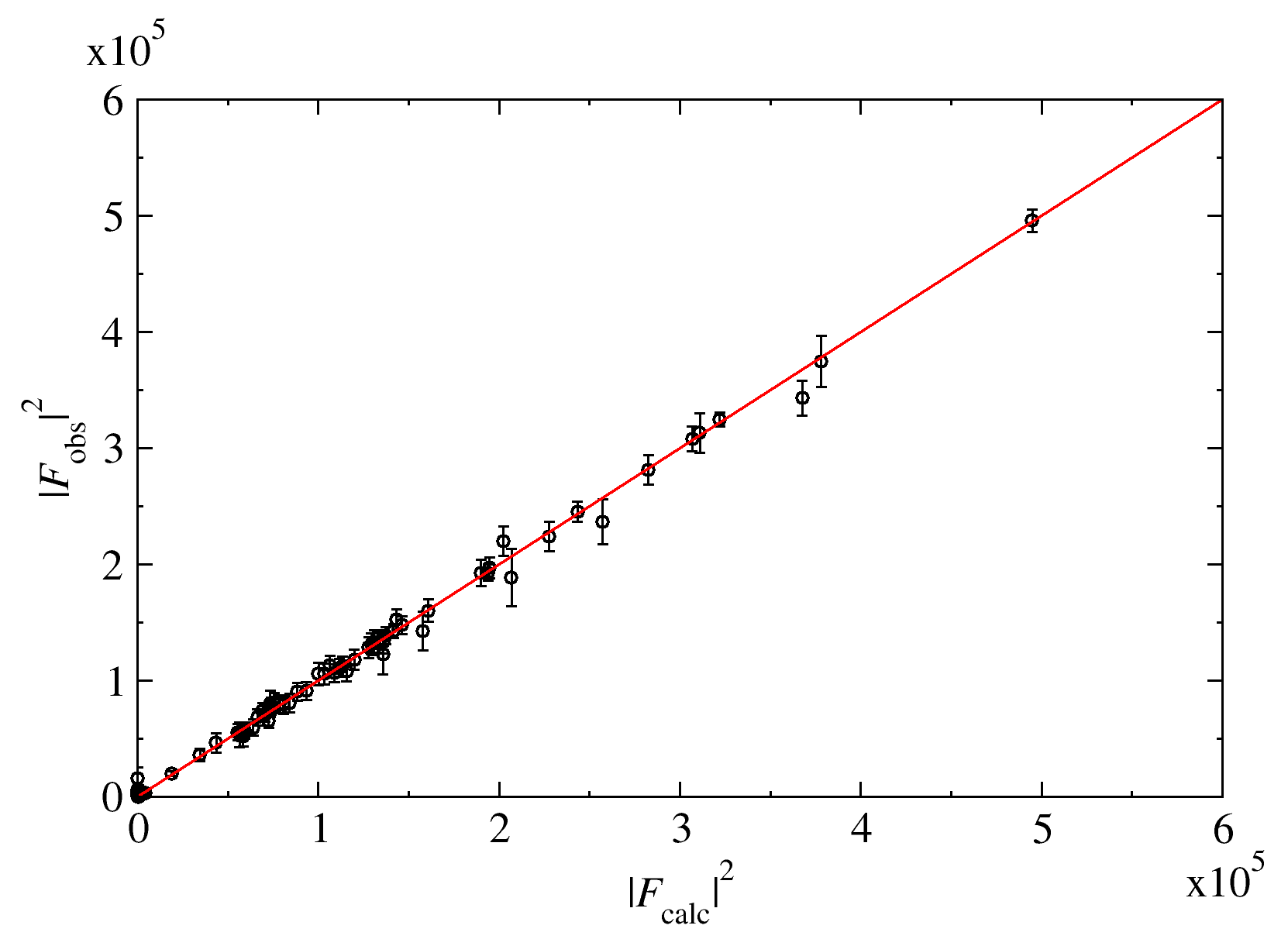}
\caption{\label{FIG::Fcomp_triangular}Lorentz-corrected x-ray diffraction intensities, $|F_\mathrm{obs}|^2$, plotted against calculated values, $|F_\mathrm{calc}|^2$, from the Rietveld refinement of the Type I triangular structure. The solid red line indicates 1:1 agreement between $|F_\mathrm{obs}|^2$ and $|F_\mathrm{calc}|^2$.}
\end{figure}

\begin{figure}
\includegraphics[width=8.5cm]{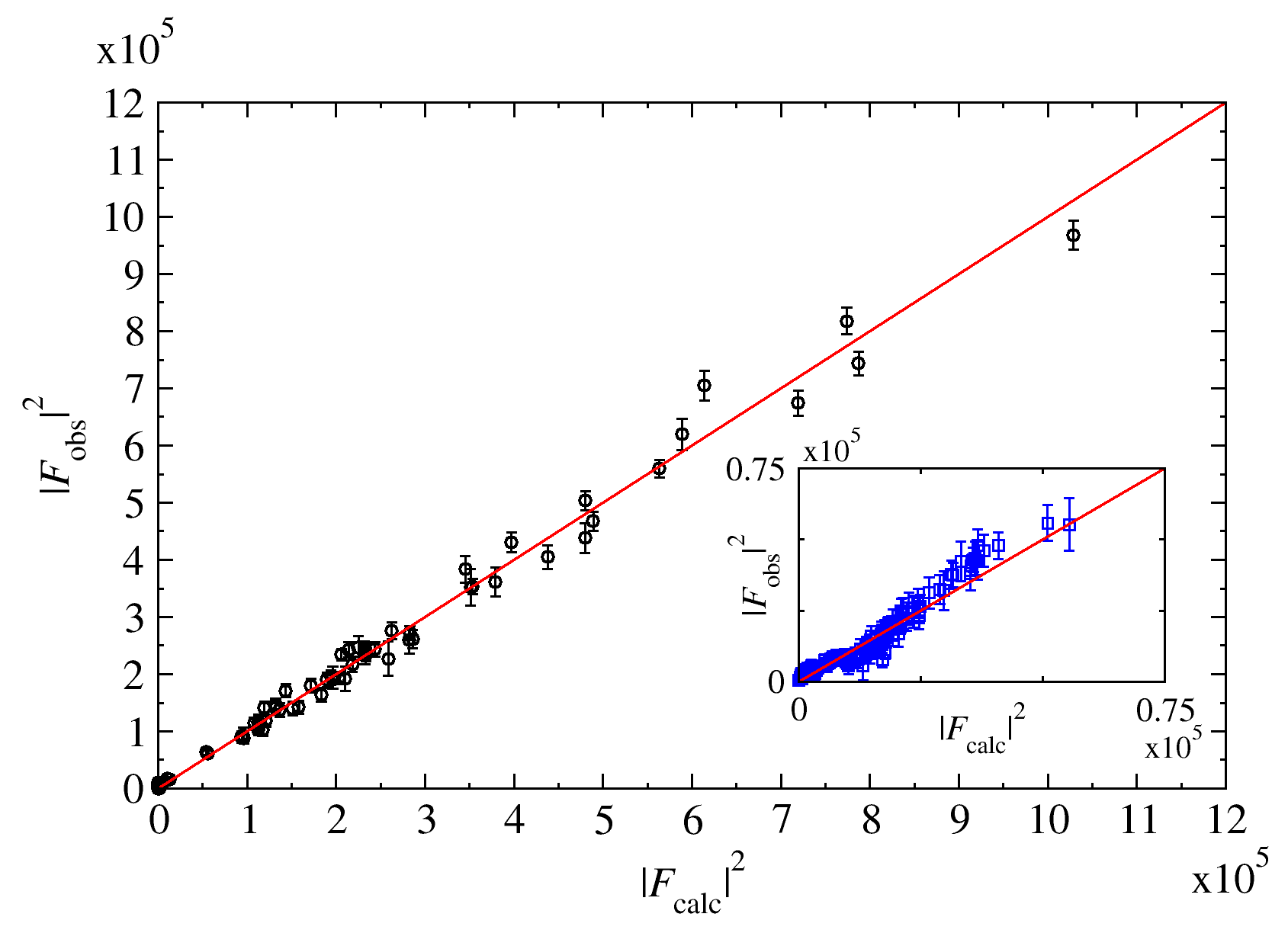}
\caption{\label{FIG::Fcomp_honeycomb}Lorentz-corrected x-ray diffraction intensities, $|F_\mathrm{obs}|^2$, plotted against calculated values, $|F_\mathrm{calc}|^2$, from the Rietveld refinement of the Type II honeycomb structure. Parent intensities are shown as black circles in the main pane, and the triple-cell super structure intensities are shown as blue squares in the inset. The solid red line indicates 1:1 agreement between $|F_\mathrm{obs}|^2$ and $|F_\mathrm{calc}|^2$.}
\end{figure}

In this Appendix we show the measured, Lorentz-corrected x-ray diffraction intensities, $|F_\mathrm{obs}|^2$, plotted against calculated values from the Rietveld refinement of the Type I (Figure \ref{FIG::Fcomp_triangular}) and Type II (Figure \ref{FIG::Fcomp_honeycomb}) structural models described in the main text. In both figures the `parent' diffraction intensities are plotted as black circles, and the triple-cell peaks of the Type II honeycomb structure are shown in the inset to Figure \ref{FIG::Fcomp_honeycomb} (blue squares). The average amplitude (square root of $|F_\mathrm{obs}|^2$) of the triple cell peaks is approximately 30\% of the average parent peak amplitude, indicating that their origin lies in a super-structure modulation of the strongly scattering iridium sublattice.

\newpage
\section{\label{SEC::k2ir03_structure}Theoretical K$_2$IrO$_3$ crystal structure}

In this Appendix we tabulate crystal structure parameters for the theoretically proposed end-member composition K$_2$IrO$_3$. Note that the previously named `Ir3' site is relabelled `K3' as for this structure it is solely occupied by potassium.

\begin{table}
\caption{\label{TAB::k2ir03_structure} Parameters of the theoretically predicted K$_2$IrO$_3$ honeycomb structure.}
\begin{ruledtabular}
\begin{tabular}{c l c c c c c}
\multicolumn{6}{l}{\textbf{Cell parameters}} \\
\multicolumn{6}{l}{Space group: $P6_322$ (\#182)} \\
\multicolumn{6}{l}{$a',b',c'$ ($\mathrm{\AA}$): 5.282, 5.282, 13.544} \\
\multicolumn{6}{l}{Volume ($\mathrm{\AA}^3$) 327.25}\\
\\
\multicolumn{6}{l}{\textbf{Atomic fractional coordinates}} \\
Atom & Site & $a$ & $b$ & $c$ & Occ.\\
\hline
Ir1 & $2b$ & 0 & 0 & 1/4 & 1 \\
Ir2 & $2c$ & 1/3 & 2/3 & 1/4 & 1 \\
K3(Ir3) & $2d$ & 2/3 & 1/3 & 1/4 & 1 \\
K2 & $6g$ & 0 & 0.34107 & 0 & 1 \\
O  & $12i$ & 0.33275 & 0.26870 & 0.15667 & 1 \\
\\
\multicolumn{6}{l}{\textbf{Selected bond lengths} ($\mathrm{\AA}$)} \\
\multicolumn{4}{l}{K2 - O $\quad$ 2.5654, 2.8580, 2.9000} & \multicolumn{2}{c}{Ir1 - O $\quad$ 2.0511} \\
\multicolumn{4}{l}{K3 - O $\quad$ 2.4515} & \multicolumn{2}{c}{Ir2 - O $\quad$ 2.0550} \\
\\
\end{tabular}
\end{ruledtabular}
\end{table}

\end{document}